\documentclass[english,10pt]{article}
\PassOptionsToPackage{hyphens}{url}
\usepackage{float}
\usepackage{subcaption}
\usepackage{fancyhdr}  
\usepackage{amsfonts}
\usepackage{tabularx}
\usepackage{booktabs}
\usepackage{lastpage}  

\usepackage{geometry}
\usepackage[sorting=none]{biblatex}

\usepackage[utf8]{inputenc}
\usepackage{physics,amssymb}  

\usepackage{comment}
\usepackage{siunitx}
\usepackage{adjustbox}

\usepackage{csquotes}
\AtBeginDocument{\RenewCommandCopy\qty\SI}
\usepackage{import}

\usepackage{pgfplots}
\DeclareUnicodeCharacter{2212}{−}
\DeclareUnicodeCharacter{2265}{+}
\usepgfplotslibrary{groupplots,dateplot}
\usetikzlibrary{patterns,shapes.arrows}
\pgfplotsset{compat=newest}

\usepackage{xcolor}     
\usepackage{graphicx}
\usepackage{multirow}
\usepackage{hyperref}         
\usepackage{listings}
\usepackage{algorithm}
\usepackage[noend]{algpseudocode}

\usepackage{todonotes}
\usepackage{tikz}
\usepackage{tabularray}

\usepackage{mathdots}
\usepackage[title]{appendix}
\usetikzlibrary{quantikz}
\hypersetup{ 
    colorlinks,
    linkcolor={red!50!black},
    citecolor={blue!50!black},
    urlcolor={blue!80!black}}

\usepackage{amsmath}

\begin{document}

\title{Toward Reliable Patient-Specific Aortic Strain Mapping from 4D CTA: Validation, Spectral Structure, and Clinical Potential}
\author{Tricia Lobo\textsuperscript{1,2,*}, Wensi Wu$^{3,4,5}$, Harold I. Litt$^{2,6}$, Nimesh D. Desai$^{7}$, Alison M. Pouch $^{2,8,\dagger}$} 
\date{\today}

\maketitle

\begin{minipage}[t]{14cm}
    \small
    $^1$  Penn Institute for Computational Science, University of Pennsylvania, Philadelphia, PA.
    
    $^2$ Department of Radiology, University of Pennsylvania, Philadelphia, PA.
    
    $^3$ Department of Mechanical Engineering and Applied Mechanics, University of Pennsylvania, Philadelphia, PA.

    $^4$ Department of Pediatrics, Division of Cardiology, Children’s Hospital of Philadelphia, Philadelphia, PA.
        
    $^5$ Cardiovascular Institute, Children’s Hospital of Philadelphia, Philadelphia, PA.

    $^6$ Diagnostics Institute, Cleveland Clinic Florida, Weston, FL.
    
    $^7$ Department of Surgery, University of Pennsylvania, Philadelphia, PA.
    
    $^8$ Department of Bioengineering, University of Pennsylvania, Philadelphia, PA.

\medskip
\noindent
\noindent\raggedright
$^{*}$Corresponding author: Tricia Lobo\\
\hspace*{1.5em}\texttt{tricia.lobo@pennmedicine.upenn.edu};
\texttt{tricia.lobo@alumni.upenn.edu}\\
$^{\dagger}$Corresponding author: Alison M. Pouch\\
\hspace*{1.5em}\texttt{pouch@pennmedicine.upenn.edu}
\end{minipage}

\label{sec:abstract}
\bigskip
\noindent\textbf{Objective} To develop and validate a framework for spatially resolved ascending aortic strain estimation from 4D computed tomography angiography (CTA) and characterize the reliability of strain measurements across segmentation sources and spatial scales.

\noindent\textbf{Methods} We constructed patient-specific aortic strain maps using nnU-Net segmentation, optimized surface remeshing, deformable registration, and mesh-based strain estimation. Finite element (FE)-derived synthetic 4D CT sequences provided controlled reference deformations for deformable registration parameter selection and validation. Strain reliability was characterized across segmentation sources and Laplace-Beltrami (LB) spectral scales. Exploratory associations with diameter, aortic height index (AHI), and age were assessed, and strain maps were qualitatively examined in three patients with high-grade aortic regurgitation (AR). 

\noindent\textbf{Results} Registration-derived median strain showed minimal bias relative to FE-reference strain for both observer-derived (bias = $0.0013$) and nnU-Net-derived meshes (bias = $0.0056$). Median strain exhibited low bias and narrow 95\% limits of agreement (LoA) across segmentation sources (interobserver: bias = $-0.0045$, LoA $[-0.0490, \,0.0399]$; nnU-Net versus observer: bias = $-0.0052$, LoA $[-0.0340, \,0.0237]$). Differences were greater for 95th-percentile strain, particularly in the interobserver comparison. LB spectral analysis showed preferential preservation of low-frequency strain organization, with $k=5$ reconstructions reducing segmentation-related disagreement by $52-56\%$. The 95th-percentile areal strain was inversely associated with maximum ascending aortic diameter ($r = -0.493$, $p = 0.005$), AHI ($r = -0.488$, $p = 0.005$), and age ($r = -0.479$, $p = 0.006$). Qualitative examination of three high-grade AR cases revealed heterogeneous regional strain patterns.

\noindent\textbf{Conclusion} This framework reproducibly estimates ascending aortic strain from 4D CTA, characterizes its reliability across segmentation sources and spatial scales, and  demonstrates its potential for noninvasive assessment of patient-specific aortic mechanics.


\section{Introduction}
\label{sec:intro}
Clinical use of image-derived strain currently centers on left ventricular global longitudinal strain (LVGLS), which can be derived through speckle-tracking echocardiography \cite{voigt_2015, potter2018assessment} or cardiovascular magnetic resonance (CMR) \cite{myocardial_strain_imaging}. Because LVGLS is sensitive to subtle systolic dysfunction, it has become an important marker of early myocardial impairment, including in heart failure with preserved ejection fraction, valvular heart disease, and cardio-oncology surveillance for subclinical treatment-related cardiotoxicity \cite{thomas_clinical_2025}. The clinical success of LVGLS has motivated exploration of image-based strain in other cardiovascular structures, including cardiac valves, atrial tissue, and the aorta, where deformation measurements may provide information beyond conventional morphologic assessment \cite{nakhaei_2026, wu2026geometricfeaturetrackingapproach, ben_zekry_patient-specific_2016, el_tallawi_valve_2021}. For the aorta, diameter-based indices such as aortic height index (AHI) and aortic size index are used to assess risk of aneurysm progression and dissection \cite{isselbacher_2022_2022}. However, many patients experience adverse aortic events despite diameters below accepted intervention thresholds \cite{pape2007aortic}, motivating investigation of complementary markers that more directly characterize regional biomechanical alterations within the aortic wall.

Bicuspid aortic valve (BAV) disease provides a clinically important example of this limitation. BAV is the most common congenital heart valve defect and is characterized by two functional cusps rather than three, with morphology varying according to the pattern of cusp fusion \cite{michelena_international_2021}. In BAV, diameter-based indices do not fully capture patient-specific differences in hemodynamics and tissue remodeling; for instance, different cusp fusion patterns have been associated with distinct eccentric outflow jet patterns, wall shear stress (WSS) distributions, and aortopathy phenotypes \cite{rodriguez-palomares_mechanisms_2023, mahadevia_bicuspid_2014}. Ex vivo studies similarly suggest regional microstructural and mechanical heterogeneity not fully reflected by diameter alone \cite{xuan_regional_2023}. 

While aortic wall stress has been linked to dissection and mortality risk, stress cannot be measured directly in vivo and must be inferred under constitutive assumptions \cite{zamirpour_temporal_2024}. Aortic strain, in contrast, can be derived from image-based deformation and has been associated with aortic diameter, age-related stiffening, and vascular function \cite{bieseviciene_two-dimensional_2017, h_evaluation_2022, redheuil_reduced_2010, vander_linden_stiffness_2023, bu_ascending_2018}. However, before image-derived strain can be incorporated into clinical or research workflows, its reliability must be characterized across segmentation sources, controlled reference deformations, and spatial scales. Further work is also needed to determine how image-derived strain relates to clinically meaningful aspects of aortic disease.

Sources of strain variability are well-documented and include imaging modality, spatial and temporal image resolution, software-related factors such as smoothing and strain formulation (Lagrangian vs. Eulerian), and operator-related factors \cite{amzulescu_2019}. Despite these sources of variability, LVGLS has gained clinical traction because it provides a reproducible global measure of myocardial deformation \cite{karlsen_global_2019, gong_reliability_2025}, with prognostic value beyond conventional functional indices \cite{kalam_prognostic_2014, kim_clinical_2025, kazukauskiene_left_2021}. In contrast, local strain estimates remain more susceptible to smoothing and exhibit greater observer- and vendor-dependent variability \cite{thomas_clinical_2025, mirea_variability_2018}. These observations motivate characterization of ascending aortic strain reliability for both global summaries and localized strain features.

Prior computational validation work on image-derived strain has largely focused on two related questions. One line of work has investigated the fundamental limits of strain estimation imposed by image quality and registration uncertainty, including theoretical and empirical analyses based on the Cramér–Rao lower bound \cite{ketcha_2017}. Another line of inquiry has evaluated whether registration-derived strain can reproduce synthetic or finite element (FE)-derived ground truth deformation fields under controlled conditions \cite{aggarwal_strain_2023, dong_measurement_2024}. However, less is known about how segmentation-source variability affects ascending aortic strain reliability across spatial scales, or whether patient-specific aortic strain reflects clinically relevant remodeling features.

Using time-resolved 4D computed tomography angiography (CTA) as a high-resolution basis for patient-specific aortic geometry and deformation tracking, we make three key contributions to address this gap:

\begin{enumerate}
    \item We present a predominantly automated framework for estimating patient-specific aortic areal strain maps, integrating segmentation, meshing, deformable registration (DR), and mesh-based strain computation in a reproducible workflow.
   \item We establish a FE-informed, strain-based validation framework for tuning DR regularization parameters by directly comparing registration-derived areal strain with known reference strain fields from patient-specific FE simulations.
    \item We characterize global strain reproducibility across segmentation sources, evaluate strain reliability across spatial scales, and investigate associations between registration-derived strain and established clinical measures of aortic remodeling.
\end{enumerate}

Although demonstrated using CT, the proposed framework for deformation tracking, strain estimation, and reliability assessment may be adaptable to other time-resolved imaging modalities.

\section{Related Work}
\label{sec:related_work}
\subsection{Accuracy and reliability of spatially resolved image-derived strain}

Existing studies demonstrate the feasibility and potential clinical relevance of image-derived aortic strain \cite{catala-santarrufina_automatic_2025, bosshardt_three-dimensional_2025}. Recently, Bosshardt et al. combined deep-learning-based aortic segmentation with nonrigid registration of free-breathing four-dimensional CMR to quantify three-dimensional distensibility, longitudinal strain, and displacement in patients with Marfan syndrome (MFS), finding significantly reduced distensibility and displacement in patients with MFS compared with healthy volunteers \cite{bosshardt_three-dimensional_2025}. However, this work does not establish the numerical accuracy or spatial reliability of dense, three-dimensional surface-resolved strain fields. More broadly, prior cardiovascular image-analysis work has demonstrated that uncertainty in image-derived segmentation can propagate to downstream quantitative model outputs \cite{sankaran_fast_2015}. These measurements introduce validation challenges that are not captured by global or segment-averaged measures alone: errors may arise from segmentation, surface correspondence, registration, and strain computation, and may affect fine-scale regional features more strongly than broad spatial patterns.

Prior aortic-strain studies have primarily evaluated reproducibility for global or cross-sectional measurements. Teixeira et al. demonstrated excellent interobserver reproducibility for two-dimensional speckle-tracking-derived global circumferential ascending-aortic strain, with minimal bias and narrow limits of agreement (LoA) \cite{teixeira_circumferential_2013}. Multicenter magnetic resonance imaging (MRI) studies have likewise reported low intraobserver and interobserver variability for aortic strain and pulse-wave-velocity measurements \cite{hrabak-paar_variability_2020}. However, the reproducibility of spatially resolved regional strain patterns remains less well-characterized.

Phantom experiments have enabled controlled evaluation of image-derived strain against sparse physical measurements. Early cine phase-contrast MRI (PC-MRI) studies evaluated vessel-wall displacement against sonomicrometry measurements in a deformable vessel phantom, demonstrating close correspondence between MRI-derived and independently measured displacement \cite{wedding_measurement_2002}. More recent phantom studies used displacement encoding with stimulated echoes (DENSE) MRI to estimate circumferential strain and compared these estimates with strain calculated from tracked embedded wire markers, showing that regional variation in circumferential strain could be recovered despite systematic bias in absolute strain magnitude \cite{wilson_vitro_2022}. However, these approaches provide reference measurements only at discrete locations or regions, rather than spatially dense ground truth.

Full-field optical methods provide an alternative approach for spatially dense experimental validation. Stereo digital image correlation (stereo-DIC) estimates three-dimensional surface displacement and strain by tracking surface texture across synchronized camera views and has been applied to human cardiovascular tissue both ex vivo and intraoperatively \cite{hokka_vivo_2015}. Separately, intraoperative optical measurement of human ascending aortic deformation using sparse sutured markers has also been demonstrated \cite{parikh_intra-operative_2021}. Unlike sparse marker-based measurements, stereo-DIC can provide distributed reference measurements over the optically accessible surface, although these measurements remain experimentally derived and limited to the observed region.

Synthetic deformation offers a computational means of validating image-derived strain against known ground truth. Prescribed affine deformations have been used to evaluate registration-derived strain accuracy \cite{aggarwal_strain_2023}, while synthetic image sequences with FE-prescribed deformation have been used to validate registration-derived aortic strain against known displacement and strain fields \cite{dong_measurement_2024, jamshidian_4d-cta_2025}, including under controlled variations in loading conditions, tissue motion, signal intensity, and image noise \cite{dong_measurement_2024}. Barbaroux et al generated realistic cine DENSE images with known high-resolution strain by simulating key features of the in vivo acquisition process \cite{barbaroux_dense-sim_2025}. Relatedly, Mukherjee et al. developed an FE-derived in-silico heart phantom to benchmark registration-based LV strain estimation \cite{mukherjee_-silico_2024}. To our knowledge, synthetic deformation has not yet been used to validate heterogeneous, spatially resolved 3D strain over human-derived ascending-aortic surfaces.

Prior work has shown that aortic strain estimates can be sensitive to segmentation variability, with pixel-scale boundary changes producing substantial differences in computed circumferential strain \cite{wilson_vitro_2022}. Controlled perturbations of aortic wall geometry have similarly been shown to alter both the magnitude and spatial distribution of registration-derived strain, with peak strain exhibiting greater sensitivity than the 99th-percentile strain \cite{miller_impact_2026}. Averaging motion estimates derived from repeated segmentations has been used to reduce random variability, damping local strain maxima while preserving broader spatial patterns \cite{hegner_using_2023}. These findings suggest that localized strain features may be more sensitive to segmentation variability than broader spatial patterns, motivating evaluation of reliability across spatial scales rather than through global metrics alone. We therefore use Laplace–Beltrami (LB) spectral analysis to quantify how strain-field reproducibility changes from large-scale structure to fine-scale variation. 

LB eigenfunctions provide an intrinsic Fourier-like basis for functions defined on curved surfaces, with low-order modes capturing slowly varying global structure and higher-order modes representing progressively finer spatial features \cite{vallet_spectral_2008}. In medical image analysis, the LB spectrum has been used for anatomical shape characterization and comparison    \cite{niethammer_global_2007}. Tan and Qiu developed LB spectral wavelets to characterize spatially localized components of functions defined on anatomical surfaces and applied the approach to hippocampal shape deformation \cite{tan_spectral_2015}. Similarly, Germanaud et al. decomposed cortical curvature into LB spectral bands and demonstrated that different frequency ranges corresponded to distinct anatomical scales of cortical folding \cite{germanaud_larger_2012}. Manifold-harmonic representations have also been used to reconstruct electroanatomical maps and estimate the spatial sampling density required to preserve specified reconstruction fidelity \cite{sanroman-junquera_arrhythmia_2018}. Related biomechanical work has used manifold harmonic filtering of image-derived valve geometry to systematically remove fine-scale surface features and evaluate their influence on FE-predicted valve mechanics \cite{watson_integrated_2017}. However, we identified no prior study using progressive LB reconstruction to quantify the reliability of a dense image-derived biomechanical field as a function of intrinsic spatial scale. We therefore use LB reconstruction to determine whether segmentation-dependent disagreement in aortic strain maps primarily reflects broad spatial organization or finer localized structure.

Here, we adopt a validation framework in which each component of the pipeline is evaluated using evidence appropriate to its claim. Segmentation agreement evaluates geometric consistency; controlled FE deformation evaluates recovery of known displacement and strain; comparisons across segmentation sources evaluate how segmentation variability affects downstream surface correspondence and strain estimates; and spectral analysis evaluates whether reliability differs between large-scale structure and finer spatial variation. Together, these evaluations characterize the accuracy and reliability of the estimated strain fields more directly than either a single global metric or a sparse local measurement.

\subsection{Deformable registration for cardiovascular motion estimation}

DR has been widely used to estimate dense spatial transformations between images, enabling characterization of tissue motion \cite{sotiras_deformable_2013}. Conventional approaches estimate a deformation field by iteratively optimizing an objective that combines an image-similarity term with spatial regularization \cite{sotiras_deformable_2013}. The resulting deformation fields can then be applied to segmented surfaces, enabling estimation of displacement and strain. More recently, learning-based registration methods have enabled substantially faster inference \cite{balakrishnan_voxelmorph_2019} and have been applied to cardiovascular motion estimation from CMR \cite{movahed_cardiomorphnet_2026}, cardiac CT \cite{yan_motion_compensated_2024}, and echocardiography \cite{zhu_neural_2019}.

For thin vascular walls, DR can exploit image information from the vessel lumen and surrounding tissues to constrain motion estimation, rather than relying solely on the weak or partially resolved wall signal \cite{dong_measurement_2024}. Registration-based approaches have been used to estimate aortic deformation across the cardiac cycle \cite{aggarwal_strain_2023, kim_three-dimensional_2023} and over longitudinal follow-up
\cite{stoecker_local_2021}. However, these studies primarily characterize displacement or summarize deformation rather than systematically evaluating the reliability of dense three-dimensional ascending-aortic strain fields or the sensitivity of strain estimates to DR parameters.

While strain accuracy depends in part on the registration configuration used to estimate deformation, prior studies have evaluated strain recovery from synthetic or simulated cardiovascular motion using fixed registration configurations without reporting a rationale for parameter selection \cite{dong_measurement_2024}, or have reported fixed parameters without examining their effects on the resulting strain fields \cite{aggarwal_strain_2023}. Zhong et al. systematically characterized registration-parameter effects, but assessed performance using displacement accuracy rather than strain accuracy \cite{zhong_analysis_2010}. To our knowledge, registration parameters have not been systematically evaluated and selected with respect to spatial strain accuracy. Here, we use controlled FE deformation to evaluate registration configurations against known strain and select parameters in a subset of patients.

\subsection{Automated aortic segmentation}

Deep-learning methods have substantially advanced automated aortic segmentation, with convolutional neural networks frequently achieving Dice similarity coefficients above 0.90 on CTA datasets \cite{feiger_evaluation_2021}. Public datasets including ImageTBAD \cite{yao_imagetbad_2021}, AVT \cite{radl_avt_2022}, and AortaSeg24 \cite{aortaseg_24} have supported method development for aortic dissection, whole-vessel-tree segmentation, and multiclass aortic labeling, respectively. Transformer-based architectures and large-scale foundation models have further expanded automated anatomical segmentation; however, performance remains dependent on pathology and task definition, with foundation models trained predominantly on normal anatomy showing reduced robustness in aneurysmal and dissected aortas \cite{el_rahal_totalsegmentator-ai_2026}.

Despite these advances, conventional overlap metrics do not fully characterize boundary accuracy \cite{reinke_understanding_2024}, and neither overlap nor surface-distance metrics characterize how segmentation variability propagates to downstream motion estimation and biomechanical mapping. We address this gap by evaluating strain reliability across observer-derived and deep-learning-derived segmentations and by assessing the reproducibility of strain patterns across spatial scales.

\section{Methods}
\label{sec:methods}
\subsection{Study population and CTA acquisition}
This study used cardiac CTA data from a prospective BAV cohort described by Pouch et al. and analyzed previously in related work by our group \cite{pouch_ct-based_2026, guo_towards_2025}. A total of 33 adults were enrolled in an IRB-approved prospective study at the University of Pennsylvania. Eligibility criteria included suspected or confirmed BAV morphology on transthoracic echocardiography (TTE) and any degree of AR. Exclusion criteria included prior aortic valve surgery and greater than mild aortic calcification.

Of the initial 33 patients, two had only 3D CT imaging and were excluded. 14 of the 31 eligible 4D CTA cases were selected as the primary cohort, with balanced representation across sex, age, and the major morphological fusion patterns to evaluate methodological performance across phenotypic variability rather than to reproduce the population prevalence of BAV fusion patterns. The remaining 17 cases were reserved for segmentation model development and evaluation. Exploratory clinical association analyses used strain measurements from the full eligible cohort. Characteristics of the 14-patient primary cohort are summarized in Table~\ref{tab:primary_cohort_characteristics}; characteristics of the full eligible cohort are provided in the Appendix.

Retrospectively ECG-gated cardiac CTA was performed on energy-integrating (EID) and photon counting (PC) CT scanners with 6 cm detectors and 66 msec temporal resolution (SOMATOM Force and NAEOTOM Alpha, Siemens Healthineers, Malvern PA). A modified TAVR protocol was used to optimize image quality of the aortic valve apparatus at systole and diastole while maintaining radiation dose comparable to that of standard cardiac CTA. Images were reconstructed at 10-20 phases across the cardiac cycle, with slice thickness from 0.2 to 0.5 mm, and in-plane voxel spacing ranging from 0.29 to 0.63 mm. Images were
exported to DICOM files and de-identified prior to analysis.

Annual follow-up imaging, when available, was performed as part of routine clinical care and was not used in the present analysis.

\begin{table}[htbp]
\centering
\caption{Primary Cohort Characteristics}
\label{tab:primary_cohort_characteristics}
\centering
\begin{tabular}{l c}
\hline
\textbf{Characteristic} & \textbf{Value} \\
\hline
Total patients, n & 14 \\
Age, y & 57.5 (42.0--61.0) \\
Female sex, \% & 28.6 \\
Height, m & 1.73 (1.65--1.82) \\
Body mass index, kg/m$^2$ & 27.2 (25.2--30.1) \\
Ascending aortic diameter, cm & 4.4 (4.0--4.5) \\
Height-indexed aortic diameter, cm/m & 2.4 (2.2--2.6) \\
Left ventricular ejection fraction, \% & 58 (55--60) \\
$>$ Mild aortic regurgitation, \% & 57.1 \\
Follow-up visits, n & 0.5 (0--2) \\
Patients with $\geq$1 follow-up, \% & 50.0 \\
\multicolumn{2}{l}{Aortic valve morphology, \%} \\
\quad L/R fusion & 42.9 \\
\quad R/N fusion & 42.9 \\
\quad Trileaflet & 14.3 \\
\hline
\end{tabular}
\smallskip
\parbox{0.60\textwidth}{%
\footnotesize Values are reported as median (IQR), percentage, or count, as appropriate. L/R indicates left–right coronary cusp fusion; R/N, right–noncoronary cusp fusion; trileaflet denotes a tricuspid aortic valve.
}
\end{table}

\subsection{Strain quantification pipeline}
\subsubsection{Pipeline overview}
\begin{figure}[h!]
  \centering
  \includegraphics[width=\textwidth]{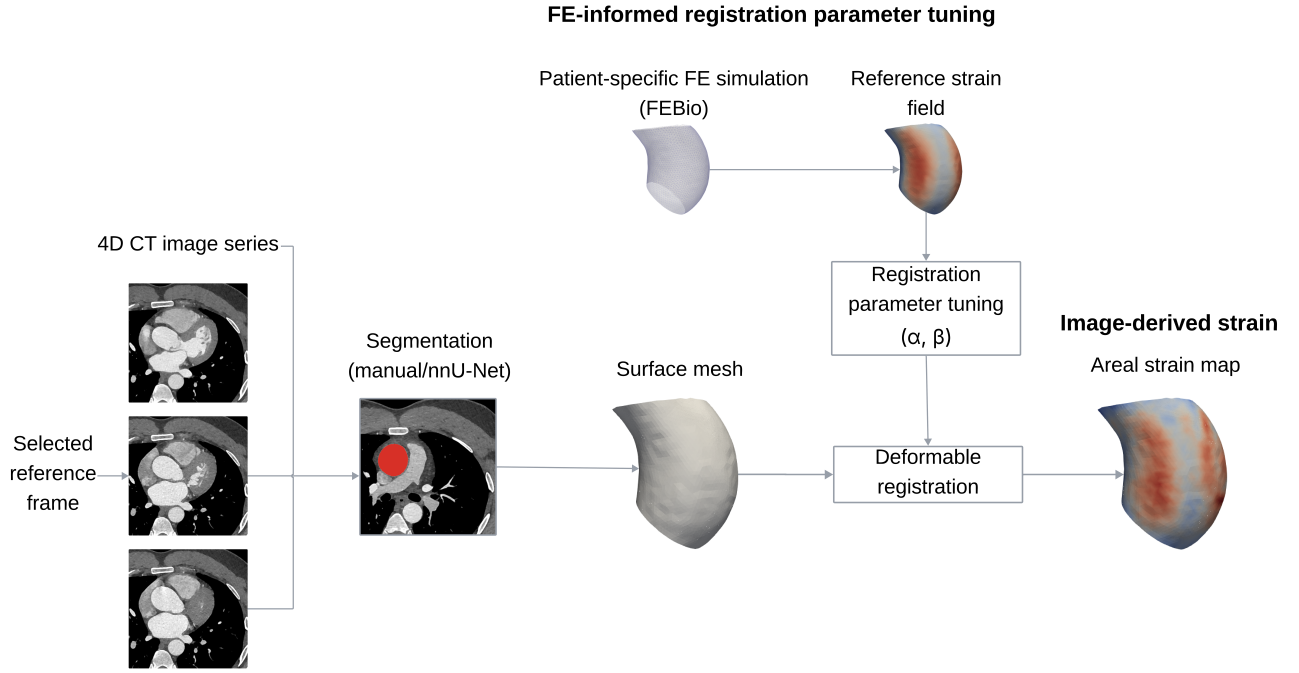}
  \caption{Overview of the image-based framework for aortic strain estimation with FE-informed registration parameter tuning. The pipeline begins with segmentation of a selected reference frame, followed by mesh generation, deformable registration, and areal strain computation. Because DR requires specification of smoothing parameters, patient-specific FE simulations were performed in a subset of five patients to identify registration parameters that best reproduced reference strain fields. The resulting parameters were then used to propagate each patient’s reference geometry across the cardiac cycle.
  }
  \label{fig:pipeline}
\end{figure}

Our strain quantification pipeline is summarized in Fig.~\ref{fig:pipeline}. Using 4D CTA images as input, the ascending aorta was segmented at a reference diastolic frame selected for minimal motion artifact. A patient-specific surface mesh was extracted from the reference segmentation and remeshed to produce a higher-quality surface mesh for DR and mesh-based strain computation.

To determine optimal DR smoothing parameters, we identified registration parameters that best reproduced reference FE displacements in a subset of five patients. For these five patients, the motion of the reference aortic geometry from the segmentation pipeline was simulated under quasi-static solid mechanics assumptions. FE-derived displacements were then used to generate synthetic time-resolved image volumes. DR was applied to the synthetic series, with metric-gradient smoothing and deformation-field smoothing parameters selected from recurring low-loss regimes identified using a composite objective incorporating geometric proximity and strain accuracy.

Subsequently for all patients, DR was used to propagate the reference geometry across the cardiac cycle, utilizing smoothing parameters derived from the FE-informed parameter tuning. This yielded displacement fields from which areal strain was computed. Details of each step are provided in subsequent sections.

\subsubsection{Image segmentation}

\paragraph{Semi-automated image segmentation}
Two observers independently performed semi-automated segmentation of the lumen of the ascending aorta using a common ITK-SNAP–based protocol \cite{py06nimg}. The first observer (Observer-1) had one year of training in cardiovascular image segmentation and analysis, while the second observer (Observer-2) was an expert in cardiovascular image analysis. Observers were blinded to each other’s segmentations. In each patient's reference image, the observers placed a user-initialized seed within the ascending aorta, followed by automated tissue classification and 3D active contour evolution. They truncated the segmentation proximally at the sinotubular junction (STJ) and extended the segmentation distally to the visible extent of the ascending aorta, with clipping at the brachiocephalic artery when present.

\paragraph{Automated segmentation using nnU-Net}
Automated aortic segmentations were generated using nnU-Net, an open-source, self-configuring framework for biomedical image segmentation \cite{isensee_nnu-net_2021}. Networks were trained using the full-resolution 3D configuration with a manually specified patch size of $192 \times 192 \times 192$ voxels. The resulting configuration used a batch size of 2, isotropic 1-mm target spacing, CT intensity normalization, and a six-stage 3D PlainConvUNet with \text{32, 64, 128, 256, 320, and 320} feature channels per stage. To promote consistent coverage of the proximal and distal extents of the ascending aorta, a custom data loader was used during training. With probability $0.8$, patch placement along the axial direction was biased toward either the proximal or distal extent of the aortic foreground label, selected with equal probability, while in-plane sampling remained unchanged. This cap-biased sampling was disabled during validation. All other training settings, including the loss function, followed the standard nnU-Net full-resolution 3D configuration.

A five-fold cross-validation scheme was employed with patient-level splitting to prevent information leakage between training and validation sets. In each fold, approximately 80\% of patients were used for training and 20\% were held out for validation. Each fold was trained for 1000 epochs. The final nnU-Net predictor was defined as the ensemble of the five fold-specific models.

\subsubsection{Mesh generation}
Surface meshes were extracted from aortic segmentations as vtkPolyData objects and cleaned to merge duplicate points and resolve non-manifold vertices. Meshes were subsequently triangulated using vtkTriangleFilter to ensure consistent element topology prior to remeshing \cite{vtkBook}. Because surface extraction does not optimize triangle quality for downstream numerical analysis, isotropic remeshing was performed to improve element shape and regularity while preserving anatomical geometry.

Surface meshes were remeshed using MMGS-O3 \cite{dobrzynski}. Remeshing parameters were optimized using Optuna with a tree-structured Parzen estimator sampler and median pruner \cite{optuna_2019}. Mesh quality was quantified using triangle aspect ratio, minimum angle, and maximum angle, which are established indicators of triangular mesh quality \cite{sorgente_survey_2023, knupp_algebraic_2003}. The optimization objective penalized elongated elements and extreme triangle angles, encouraging more regular mesh geometry for downstream strain computation. For each patient, the highest-performing Optuna trial defined the final remeshing parameters used in subsequent analyses.

\subsubsection{Deformable registration}

Registration was performed using the Greedy registration tool, an open-source implementation of regularized diffeomorphic image registration \cite{yushkevich_ic-p-174_2016, greedy_github} that has successfully been used for computing image-based strain of the aortic root in 4D transesophageal echocardiography \cite{aggarwal_strain_2023, hao_streamlining_2025}. For each image pair, Greedy estimated a smooth deformation mapping the reference image to the target by optimizing an image-similarity objective subject to regularization. The regularization parameters controlled smoothing of the image-metric gradient and deformation field and were optimized as described in Section~\ref{sec:fe_informed_parameter_selection}.

\subsubsection{Strain estimation}

Areal strain, which quantifies surface deformation via the relative change in surface area, was computed for each triangular element in the surface mesh. End diastole was defined as the reference configuration for strain computation; end-systolic strain was used for all subsequent strain analyses. Following Wu et al \cite{wu_novel_2025}, each triangular element was  parameterized using local coordinates $(\xi,\eta)$ on a reference triangle with linear shape functions,

\begin{equation}
N_1(\xi,\eta)=\xi,\qquad
N_2(\xi,\eta)=\eta,\qquad
N_3(\xi,\eta)=1-\xi-\eta,
\end{equation}
such that
\begin{equation}
\mathbf{x}(\xi,\eta)=\sum_{a=1}^{3}N_a(\xi,\eta)\mathbf{x}_a,
\end{equation}
where $\mathbf{x}_a\in\mathbb{R}^3$ denotes the Cartesian coordinates of the $a^{\mathrm{th}}$ vertex. The surface Jacobian, defined as

\begin{equation}
\mathbf{J}
=
\begin{bmatrix}
\dfrac{\partial \mathbf{x}}{\partial \xi} \\
\dfrac{\partial \mathbf{x}}{\partial \eta}
\end{bmatrix}
=
\begin{bmatrix}
\dfrac{\partial N_1}{\partial \xi} &
\dfrac{\partial N_2}{\partial \xi} &
\dfrac{\partial N_3}{\partial \xi} \\
\dfrac{\partial N_1}{\partial \eta} &
\dfrac{\partial N_2}{\partial \eta} &
\dfrac{\partial N_3}{\partial \eta}
\end{bmatrix}
\begin{bmatrix}
\mathbf{x}_1^{\mathrm{T}} \\
\mathbf{x}_2^{\mathrm{T}} \\
\mathbf{x}_3^{\mathrm{T}}
\end{bmatrix}.
\end{equation}
yields two surface tangent vectors,

\begin{equation}
\mathbf{t}_1=\frac{\partial\mathbf{x}}{\partial\xi},\qquad
\mathbf{t}_2=\frac{\partial\mathbf{x}}{\partial\eta}.
\end{equation}

For linear triangular elements, these tangent vectors are constant over each element, and the element area is computed as

\begin{equation}
A=\frac{1}{2}\left\|\mathbf{t}_1\times\mathbf{t}_2\right\|.
\end{equation}

Areal strain was then calculated as the fractional change in element area between the reference and deformed configurations,

\begin{equation}
E_{\mathrm{area}}
=
\frac{A_{\mathrm{def}}-A_{\mathrm{ref}}}{A_{\mathrm{ref}}}
\end{equation}
where $A_{\mathrm{ref}}$ and $A_{\mathrm{def}}$ denote triangular face areas in the reference end-diastolic and target end-systolic configurations.

Areal strain was computed for patient-specific meshes derived from both nnU-Net segmentations and semi-automated segmentations generated by two observers.

\subsection{Experiments}
We evaluated the proposed framework across multiple levels of the pipeline, including segmentation, surface proximity, registration and strain accuracy, spatial reliability, and clinical associations. We first assessed the effect of segmentation source on geometry by quantifying segmentation overlap and surface proximity for interobserver and nnU-Net-versus-observer comparisons; Observer-1 segmentations served as the reference for the latter. We then performed FE-informed DR parameter selection by optimizing a composite objective incorporating geometric proximity and strain accuracy. Using the selected parameters, we evaluated DR-derived strain against known FE-reference strain at the final target frame of each synthetic 4D CT sequence. Separately, we assessed the reproducibility of registration-derived strain across segmentation sources by comparing strain estimates obtained from interobserver and nnU-Net-versus-observer segmentations. Spatial robustness was assessed by reconstructing registration-derived strain fields using progressively larger sets of LB modes and quantifying cross-segmentation disagreement using root mean square deviation (RMSD) as higher-frequency components were restored.

We examined associations between end-systolic strain and established clinical measures of aortic remodeling. We also qualitatively examined strain heterogeneity within clinically defined AR severity groups. For these exploratory clinical analyses, the cohort was expanded beyond the primary cohort to include end-systolic strain maps from Observer-1 segmentations for the full eligible cohort.

\subsubsection{Technical validation and robustness}
\paragraph{Segmentation overlap and surface proximity}

The Dice similarity coefficient \cite{dice_measures_1945}, a commonly used overlap-based measure in medical image segmentation, was used to quantify volumetric overlap. The Dice coefficient for two corresponding segmentations A and B was computed as follows:

\begin{equation}
\mathrm{Dice}(A,B)
= \frac{2\,|A \cap B|}{|A| + |B|},
\end{equation}

where $|\cdot|$ denotes the number of voxels in the corresponding set and $A \cap B$ denotes the intersection region between segmentations. Higher Dice values indicate greater volumetric overlap.

Mean surface distance (MSD) was additionally used to quantify surface proximity between corresponding segmentation-derived or remeshed surfaces \cite{pouch_medially_2015}. Assuming that $\mathcal{S}_A$ and $\mathcal{S}_B$ denote the sets of surface points
for two segmentations $A$ and $B$, respectively, MSD was defined as
\begin{equation}
\mathrm{MSD}(\mathcal{S}_A,\mathcal{S}_B) = \frac{1}{2}\left(
  \frac{1}{|\mathcal{S}_A|}\sum_{x\in\mathcal{S}_A}\min_{y\in\mathcal{S}_B}\|x-y\|_2
+ \frac{1}{|\mathcal{S}_B|}\sum_{y\in\mathcal{S}_B}\min_{x\in\mathcal{S}_A}\|x-y\|_2
\right).
\label{eq:msd}
\end{equation}

\paragraph{FE-informed registration parameter selection}
\label{sec:fe_informed_parameter_selection}
We used patient-specific FE simulations as controlled references for selecting DR parameters and evaluating the recovery of known deformation and strain. To generate controlled synthetic deformations, FE simulations were performed for
five patients using patient-specific ascending aortic surface meshes derived
from both nnU-Net and semi-automated segmentations. Meshes were imported into
FEBio \cite{maas_febio_2012}, an open-source FE solver for biomechanical
simulation. The aortic wall was modeled using the
Holzapfel--Gasser--Ogden (HGO) constitutive law with tissue parameters adopted
from \cite{gasser_hyperelastic_2005}: ground-matrix shear modulus
\(c=\SI{7.64}{\kilo\pascal}\), fiber modulus
\(k_1=\SI{996}{\kilo\pascal}\), fiber exponential coefficient \(k_2=524.6\), fiber dispersion
\(\kappa=0.226\), and fiber mean orientation angle
\(\gamma=\ang{49.98}\). The same constitutive parameters were applied to all simulations, without prescribing a patient-specific material orientation field. Because these parameters were not fit to subject-specific ascending aortic tissue, they were treated as surrogate constitutive settings for validation rather than definitive
material properties.

Simulation settings were held constant across all cases, including bulk modulus
\(K=\SI{1500}{\kilo\pascal}\), applied pressure
\(p=\SI{-0.5}{\kilo\pascal}\), shell thickness
\(t=\SI{0.1}{\milli\metre}\), and density
\(\rho=\SI{1e-6}{\kilo\gram\per\milli\metre\cubed}\). Translational motion at
the proximal and distal cuffs was fixed, radial expansion was unconstrained, no
pre-stress was applied, and strain was computed relative to the undeformed
reference mesh.

To generate synthetic 4D image series, ten intermediate quasi-static
deformation states were selected between the undeformed and deformed FE
configurations. Mesh displacement fields were interpolated to dense
displacement fields using a Gaussian kernel and resampled onto the image grid
using cubic spline interpolation.

For registration parameter selection, we evaluated two image similarity metrics, normalized cross-correlation (NCC) and sum of squared differences (SSD), together with a range of regularization parameters, $\alpha$ and $\beta$, where $\alpha$ controls smoothing of the
image-metric gradient during optimization and $\beta$ controls smoothing of the
resulting deformation field. These parameters govern the tradeoff between sensitivity to localized deformation and robustness to image noise, segmentation irregularities, and discretization artifacts. For each patient, fifty DR trials were performed
while varying both the similarity metric and regularization parameters.
Parameter configurations were selected automatically using Optuna \cite{optuna_2019}, with a Tree-structured Parzen Estimator sampler and median pruning, by minimizing a composite objective comprising MSD between the registered and FE-reference meshes, strain root mean square error (RMSE) relative to the FE reference, and a slope penalty,

\begin{equation}
 \mathcal{L} = \mathcal{L}_{\mathrm{MSD}}
+ 0.5\,\mathcal{L}_{\mathrm{RMSE}}
+ 0.5\,\log\!\left(1+\sum_f (m_f-1)^2\right),
\end{equation}
where $m_f$ denotes the framewise slope between DR-derived and FE-reference strain. The RMSE term penalized absolute strain error, whereas the slope penalty discouraged systematic attenuation or amplification of the FE-reference strain; the logarithmic transform reduced the influence of large slope deviations on the composite objective.

Because the best trial for an individual patient may reflect local optimization
effects or noise, parameter selection was based on recurring low-loss regimes
rather than patient-specific optima. For each patient, trials were ranked by
composite loss, and the top 30\% were retained to identify recurring low-loss
parameter regimes. This threshold excluded poor-performing trials while retaining sufficient samples to characterize recurring low-loss parameter regimes across patients and segmentation sources. The tuning procedure selected the NCC similarity metric with $\alpha=3.21~\mathrm{mm}$ and $\beta=0.78~\mathrm{mm}$; these settings were fixed for all subsequent registration and strain analyses.

Using the selected registration parameters, the accuracy of DR-derived areal strain relative to the known FE-reference strain was evaluated at the final target frame of each synthetic 4D CT sequence. Metrics were computed for subject-level median and 95th-percentile areal strain and summarized across the five-case FE evaluation cohort. Strain was compared using Bland–Altman analysis and least-squares regression. Bland–Altman bias was defined as the mean signed FE–DR difference, with 95\% LoA calculated as bias $\pm 1.96$ SD.

\paragraph{Strain reproducibility across segmentation sources}
We evaluated the reproducibility of registration-derived strain across segmentation sources by summarizing areal strain at the end-systolic frame using the median and 95th percentile. Cross-segmentation agreement was assessed using Bland–Altman analysis, with linear association assessed separately using Pearson correlation.

\paragraph{Spectral analysis of strain reproducibility}
To evaluate the spatial scales at which strain patterns remained reproducible across segmentation sources, the comparison strain field was first mapped onto the corresponding reference mesh using nearest-neighbor correspondence \cite{2020SciPy-NMeth}, enabling vertex-wise comparison on a common surface. Both strain fields were then represented using the LB eigenfunctions of the reference mesh, so that spectral reconstructions were expressed in a common spatial basis.

Representing the surface domain $\mathcal{B}$ as a triangular FE mesh, 
we constructed the LB stiffness matrix $\mathbf{A}$ and mass matrix 
$\mathbf{M}$ using three-node linear triangular elements and homogeneous Neumann boundary conditions at the open surface boundaries, following the formulation of Costabal et al \cite{sahli_costabal_mathmi_2024}:

\begin{equation}
A_{ij}
=
\sum_{e=1}^{n_{\mathrm{el}}}
\int_{\mathcal{B}_e}
\nabla_{\mathcal{B}} N_i \cdot \nabla_{\mathcal{B}} N_j
\, d\mathcal{B},
\qquad
M_{ij}
=
\sum_{e=1}^{n_{\mathrm{el}}}
\int_{\mathcal{B}_e}
N_i N_j
\, d\mathcal{B},
\end{equation}
where $N_i$ are the linear FE shape functions and 
$\mathcal{B}_e$ denotes a triangular surface element. The assembled matrices 
define the generalized eigenvalue problem

\begin{equation}
\mathbf{A}\mathbf{v}_k
=
\lambda_k \mathbf{M}\mathbf{v}_k,
\end{equation}
where $\mathbf{v}_k$ contains the nodal values of the $k$th discrete 
LB eigenfunction and $\lambda_k$ is its corresponding eigenvalue.

The zero-eigenvalue LB eigenfunction is spatially constant, allowing a surface field to be separated into a constant component and spatially varying nonconstant modes \cite{anqi_qiu_smooth_2006}. For each strain map, the constant component was represented by its FE surface-area-weighted mean,

\begin{equation}
\bar{f}_{M}
=
\frac{
\mathbf{1}^{T}\mathbf{M}\mathbf{f}
}{
\mathbf{1}^{T}\mathbf{M}\mathbf{1}
},
\end{equation}
where $\mathbf{f}$ is the vector of nodal strain values and $\mathbf{1}$ is the all-ones vector. The mean-free field was then projected onto the nonconstant LB eigenfunctions using the FE mass-matrix inner product \cite{graichen_sphara_2015},

\begin{equation}
a_i
=
\boldsymbol{\phi}_i^{T}
\mathbf{M}
\left(
\mathbf{f}
-
\bar{f}_{M}\mathbf{1}
\right),
\end{equation}
where $a_i$ is the spectral coefficient associated with the $i$th nonconstant eigenfunction.

The strain field reconstructed using the first $k$ nonconstant eigenfunctions was therefore

\begin{equation}
\mathbf{f}_k
=
\bar{f}_{M}\mathbf{1}
+
\sum_{i=1}^{k}
a_i\boldsymbol{\phi}_i.
\end{equation}
Here, $k$ denotes the number of spatially varying LB modes retained in the reconstruction. Smaller values of $k$ preserve broad, smooth spatial variation, whereas progressively larger values admit increasingly fine-scale intrinsic spatial structure.
Reconstructions were evaluated at $k = 5$, $10$, $20$, $30$, and $50$. For each retained-mode count k, we computed the RMSD, $\Delta_k$, between the reconstructed strain fields over corresponding valid vertices. The RMSD was normalized by that of the corresponding full-resolution strain fields, $\Delta_{\mathrm{raw}}$, yielding the relative disagreement $\Delta_k/\Delta_{\mathrm{raw}}$.

\subsubsection{Exploratory clinical analyses}

\paragraph{Associations with aortic dimensions and age}
To evaluate whether registration-derived strain reflected clinically relevant aortic enlargement, associations between 95th-percentile end-systolic areal strain and maximum ascending aortic diameter and AHI were assessed using Pearson correlation in the full eligible cohort ($n = 31$), based on Observer-1 segmentations. Patient age was evaluated separately as an exploratory demographic association.

\paragraph{Strain-map patterns in aortic regurgitation}
AR severity was obtained from clinical echocardiography reports.
To explore whether areal strain captured heterogeneity within a clinically defined AR severity category, three high-grade AR subjects, defined as those with moderate-to-severe or severe AR, were examined qualitatively using CTA images and end-systolic strain maps. Because the 14-patient primary cohort included only two subjects with high-grade AR, one additional high-grade AR subject from the full eligible cohort was included for qualitative comparison.


\section{Results}
\label{sec:results}
\subsection{Technical validation and robustness}

\subsubsection{Segmentation overlap and surface proximity}
Segmentation overlap was high for both interobserver and Observer-1–nnU-Net comparisons: median Dice coefficients were 0.95 (IQR, 0.95–0.96) for interobserver comparisons and 0.96 (IQR, 0.94–0.97) for Observer-1 versus nnU-Net. Surface discrepancies were submillimetric but modestly larger for nnU-Net-versus-observer comparisons: segmentation MSD increased from 0.46 to 0.64 mm, and remeshed-surface MSD increased from 0.38 to 0.52 mm (Table~\ref{tab:segmentation_metrics}).

\begin{table}[!htbp]
\centering
\caption{Segmentation overlap and surface proximity for interobserver and nnU-Net-versus-observer comparisons. Values are reported as median (interquartile range).}
\label{tab:segmentation_metrics}
\begin{tabular}{lcc}
\hline
Metric & Interobserver (n=14) & Observer-1 vs. nnU-Net (n=14) \\
\midrule
Dice coefficient & 0.95 (0.95--0.96) & 0.96 (0.94--0.97) \\
\addlinespace
Segmentation MSD (mm) & 0.46 (0.43--0.72) & 0.64 (0.51--0.77) \\
\addlinespace
Remeshed-surface MSD (mm) & 0.38 (0.27--0.48) & 0.52 (0.38--0.79)  \\
\hline
\end{tabular}
\end{table}

\subsubsection{FE-informed registration parameter selection and strain evaluation}
\label{sec:finite_element_simulation}

\begin{figure}[!htbp]
    \centering
    \includegraphics[width=0.72\linewidth]{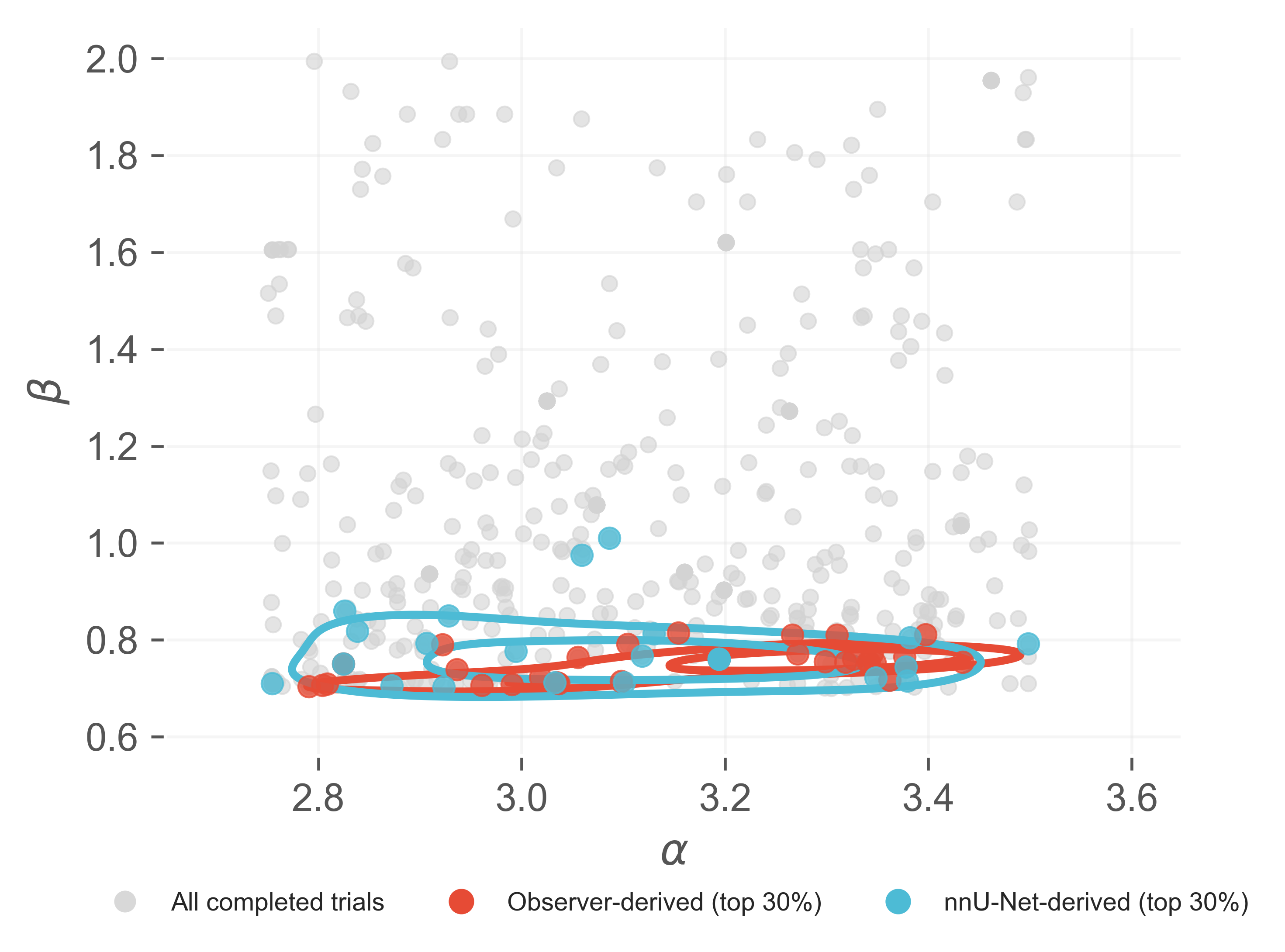}
    \caption{FE-informed registration parameter tuning. Gray points represent all completed Optuna trials. Colored points indicate NCC $3\times3\times3$ trials among the lowest-loss $30\%$ of completed trials for observer-derived and nnU-Net-derived geometries. Contours denote the $50\%$ and $80\%$ kernel-density regions of these retained NCC trials, illustrating recurring parameter regimes across segmentation sources. Here, $\alpha$ controls smoothing of the image-metric gradient during optimization, and $\beta$ controls smoothing of the resulting deformation field.}
    \label{fig:dr_parameter_tuning_ncc3x3}
\end{figure}

\paragraph{Registration parameter selection}
The retained trials, comprising the lowest-loss 30\% of completed trials within each patient and segmentation source, clustered within a common region of parameter space across segmentation sources (Fig.~\ref{fig:dr_parameter_tuning_ncc3x3}). The selected setting, $\alpha=3.21~\mathrm{mm}$ and $\beta=0.78~\mathrm{mm}$, lay within the overlapping high-density regions for both observer- and nnU-Net-derived meshes. NCC with a $3\times3\times3$ kernel was consistently represented among the retained low-loss trials across patients and segmentation sources and was therefore selected as the image-similarity metric.

\paragraph{Comparison with FE-reference strain}

\begin{figure}[htbp]
  \centering
  \includegraphics[width=0.8\textwidth]{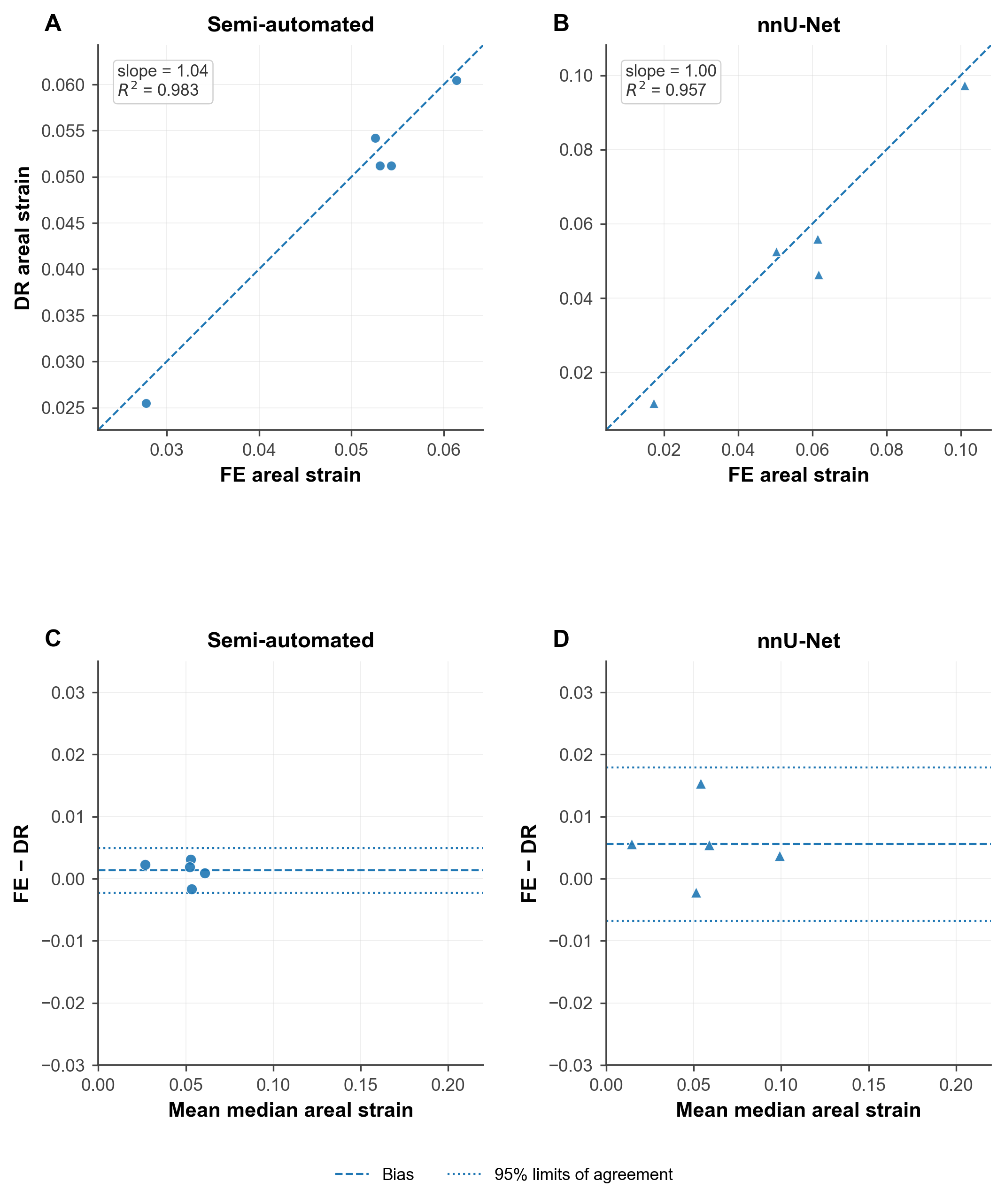}
  \caption{Validation of DR-derived areal strain against FE reference strain. (A--B) Comparison of FE-derived and DR-derived median areal strain for semi-automated and nnU-Net-derived segmentations. Each point represents one FE-validation subject. Dashed lines indicate the line of identity. (C--D) Bland--Altman analysis of median areal strain demonstrating the difference (FE--DR) as a function of the mean strain for each subject. Dashed lines denote bias and dotted lines denote the $95\%$ LoA. Quantitative comparison metrics, including the corresponding results for 95th-percentile strain, are summarized in Table~\ref{tab:fe_dr_comparison}.}
\label{fig:fe_validation_main_figure}
\end{figure}

DR-derived median areal strain showed near-one-to-one linear correspondence with FE-reference strain for both semi-automated segmentations (slope = 1.04, $R^2 = 0.983$) and nnU-Net segmentations (slope = 1.00, $R^2 = 0.957$; Fig.~\ref{fig:fe_validation_main_figure}). Bias was $0.0013$ for semi-automated segmentations and $0.0056$ for nnU-Net segmentations, with corresponding 95\% LoA of $[-0.0023,\,0.0049]$ and $[-0.0068,\,0.0179]$, respectively (Table~\ref{tab:fe_dr_comparison}). 

\begin{table}[ht!]
\centering
\caption{Comparison of registration-derived areal strain with finite-element reference strain. Bias denotes the mean signed difference between the first and second method listed in each comparison. LoA are reported as bias $\pm 1.96$ SD.}
\label{tab:fe_dr_comparison}
\begin{tabular}{llccc}
\toprule
Segmentation source & Metric & Bias & 95\% LoA & Pearson r \\
\midrule
Semiautomated & Median areal strain & 0.0013 & [-0.0023, 0.0049] & 0.991 \\
Semiautomated & 95th-percentile areal strain & 0.0108 & [0.0022, 0.0194] & 0.997 \\
nnU-Net & Median areal strain & 0.0056 & [-0.0068, 0.0179] & 0.978 \\
nnU-Net & 95th-percentile areal strain & 0.0038 & [-0.0265, 0.0340] & 0.977 \\
\bottomrule
\end{tabular}
\end{table}

\subsubsection{Strain reproducibility across segmentation sources}
\begin{figure}[ht!]
  \centering
  \includegraphics[width=0.8\textwidth]{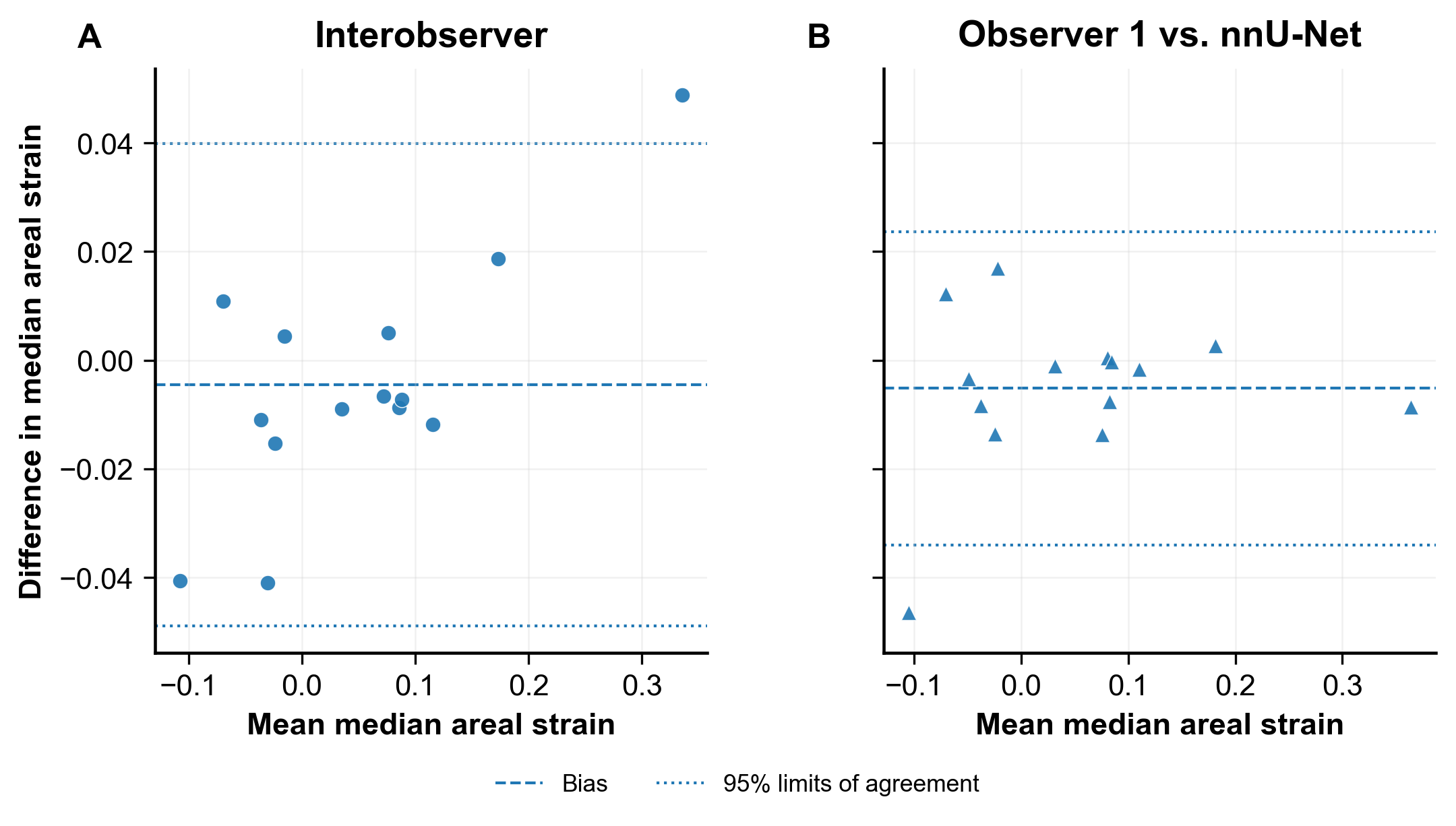}
  \caption{Strain agreement across segmentation sources. Each point represents one subject. (A) Bland--Altman analysis of median areal strain comparing Observer-1 and Observer-2. (B) Bland--Altman analysis of median areal strain comparing Observer-1 and nnU-Net. Dashed lines denote bias and dotted lines denote the $95\%$ LoA. Quantitative comparison metrics, including corresponding results for 95th-percentile strain, are summarized in Table~\ref{tab:interobserver_strain_comparison}.}
\label{fig:ba_cross_segmentation_agreement}
\end{figure}

Median areal strain showed very strong correlations for both interobserver and nnU-Net-versus-observer comparisons ($r = 0.991$ and $r = 0.993$, respectively), with low bias (-0.0045 and -0.0052, respectively) and 95\% LoA of $[-0.0490,\,0.0399]$ and $[-0.0340,\,0.0237]$, respectively
(Fig.~\ref{fig:ba_cross_segmentation_agreement}; Table~\ref{tab:interobserver_strain_comparison}). For 95th-percentile strain, correlations were $ r= 0.948$ and $r = 0.997$, with biases of -0.0371 and 0.0029, respectively, and 95\% LoA of $[-0.1810,\,0.1068]$ and $[-0.0347,\,0.0406]$.

\begin{table}[ht!]
\centering
\caption{Comparison of areal strain across interobserver and nnU-Net-versus-observer pairs. Bias denotes the mean signed difference between the first and second method listed in each comparison. LoA are reported as bias $\pm 1.96$ SD.}
\label{tab:interobserver_strain_comparison}
\begin{tabular}{llccc}
\toprule
Comparison & Metric & Bias & 95\% LoA & Pearson r \\
\midrule
Observer-1 vs. Observer-2 & Median areal strain & -0.0045 & [-0.0490, 0.0399] & 0.991 \\
Observer-1 vs. Observer-2 & 95th-percentile areal strain & -0.0371 & [-0.1810, 0.1068] & 0.948 \\
Observer-1 vs. nnU-Net & Median areal strain & -0.0052 & [-0.0340, 0.0237] & 0.993 \\
Observer-1 vs. nnU-Net & 95th-percentile areal strain & 0.0029 & [-0.0347, 0.0406] & 0.997 \\
\bottomrule
\end{tabular}
\end{table}

\subsubsection{Spectral analysis of strain map reproducibility}

\begin{figure}[htbp]
    \centering
    \includegraphics[width=1.0\linewidth]{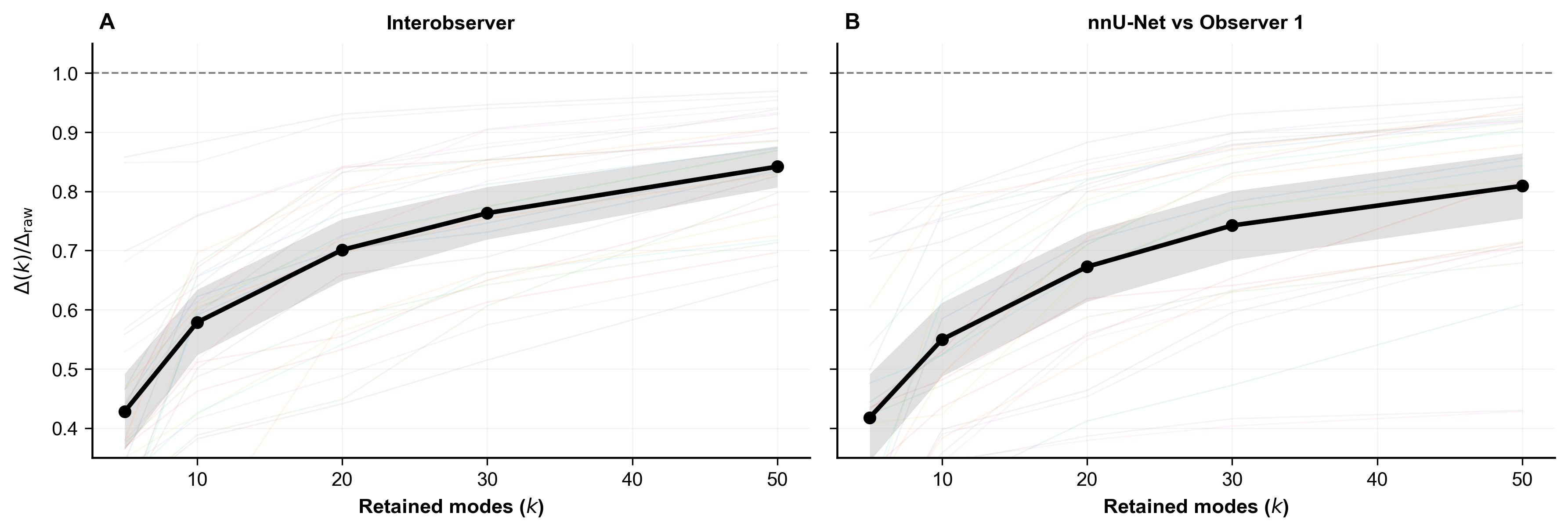}
    \caption{Relative strain disagreement retained after spectral truncation for (A) interobserver and (B) nnU-Net-versus-Observer-1 comparisons. Colored lines show individual subjects, the black line indicates the cohort mean, and the shaded region represents the 95\% confidence interval. Mean strain disagreement at each truncation level is normalized by the unreconstructed strain disagreement ($\Delta k / \Delta_{\rm raw}$). The substantially lower normalized disagreement at low retained-mode counts indicates that segmentation-dependent differences are preferentially concentrated in higher spatial-frequency components of the strain field.}
    \label{fig:relative_spectral_diff}
\end{figure}

\begin{figure}[ht!]
    \centering
\includegraphics[width=1.0\linewidth]{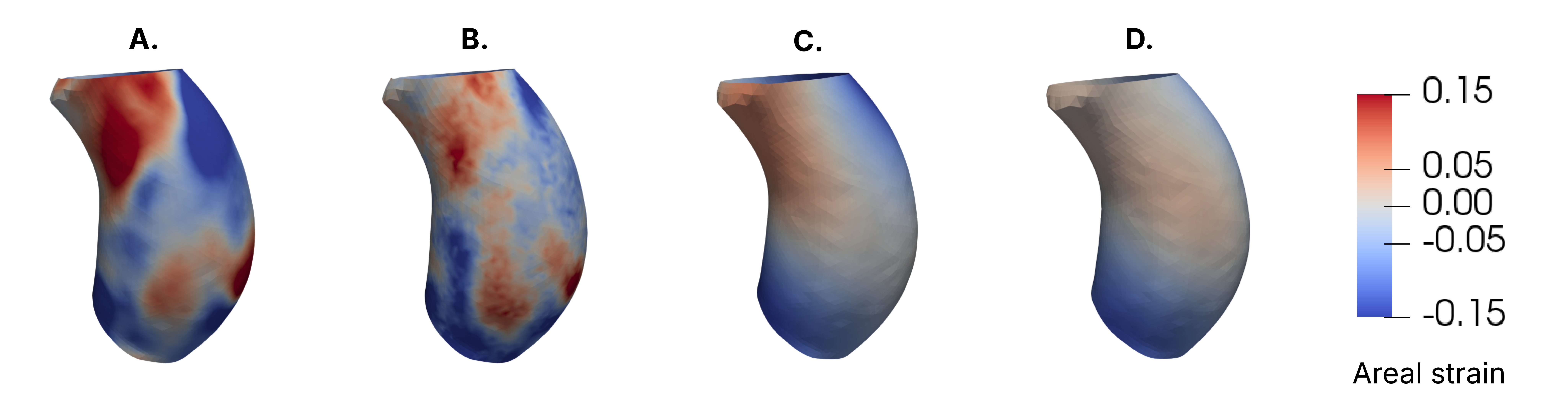}
    \caption{Representative interobserver spectral reconstruction illustrating preservation of coarse strain organization under low-frequency spectral truncation. (A) Full-resolution strain map for Observer-1. (B) Full-resolution Observer-2 strain map projected onto the Observer-1 reference mesh. (C) Observer-1 strain field reconstructed using the first $k=10$ Laplace–Beltrami eigenfunctions of the reference mesh. (D) Projected Observer-2 strain field reconstructed using the same $k=10$ eigenfunctions. $k=10$ is shown as an intermediate representative truncation level for visualization and was not selected as an optimal cutoff. Although localized hotspots visible in the full-resolution maps are attenuated under low-frequency reconstruction, the dominant coarse spatial organization is preserved across segmentation sources.}
    \label{fig:spectral_reconstruction}
\end{figure}

Across interobserver and nnU-Net–versus-observer comparisons, strain disagreement was greater at higher spatial frequencies. At $k=5$, spectral reconstruction retained approximately $44–48\%$ of the disagreement present in the full-resolution strain fields, corresponding to a $52–56\%$ reduction across reference meshes and segmentation-source comparisons (Table~\ref{tab:spectral_truncation}). Disagreement increased progressively as more higher-order modes were retained (Figure~\ref{fig:relative_spectral_diff}), while low-order reconstructions preserved the dominant spatial organization of the strain fields (Figure~\ref{fig:spectral_reconstruction}). Together, these findings indicate that broad strain patterns were more reproducible across segmentation sources than localized fine-scale variation.

\begin{table}[H]
\centering
\caption{Mean strain-field RMSD under spectral truncation. Values are reported as mean $\pm$ standard deviation across subjects for raw strain fields and reconstructions retaining $k \in \{5,10,50\}$ Laplace–Beltrami modes.}
\label{tab:spectral_truncation}
\resizebox{0.9\textwidth}{!}{
\begin{tabular}{llcccc}
\toprule
Comparison & Reference mesh & Raw & $k=5$ & $k=10$ & $k=50$ \\
\midrule
Observer-1 vs Observer-2 & Observer-1 & 0.094 ± 0.067 & 0.042 ± 0.034 & 0.056 ± 0.040 & 0.082 ± 0.063 \\
Observer-1 vs Observer-2 & Observer-2 & 0.087 ± 0.052 & 0.038 ± 0.032 & 0.050 ± 0.036 & 0.075 ± 0.051 \\
nnU-Net vs Observer-1 & Observer-1 & 0.054 ± 0.034 & 0.026 ± 0.025 & 0.033 ± 0.027 & 0.047 ± 0.033 \\
nnU-Net vs Observer-1 & nnU-Net & 0.050 ± 0.029 & 0.024 ± 0.023 & 0.032 ± 0.025 & 0.043 ± 0.029 \\
\bottomrule
\end{tabular}}
\end{table}

\subsection{Exploratory clinical analyses}

\paragraph{Associations with aortic dimensions and age}
In the full cohort, 95th-percentile areal strain was inversely associated with maximum ascending aortic diameter (Pearson $r = -0.504$, $p = 0.004$), AHI ($r = -0.476$, $p = 0.007$), and age ($r = -0.495$, $p = 0.005$). In a sensitivity analysis excluding one influential observation, all three Pearson associations were attenuated and no longer statistically significant. 

\begin{figure}[htbp]
    \centering    
    \includegraphics[width=0.7\linewidth]{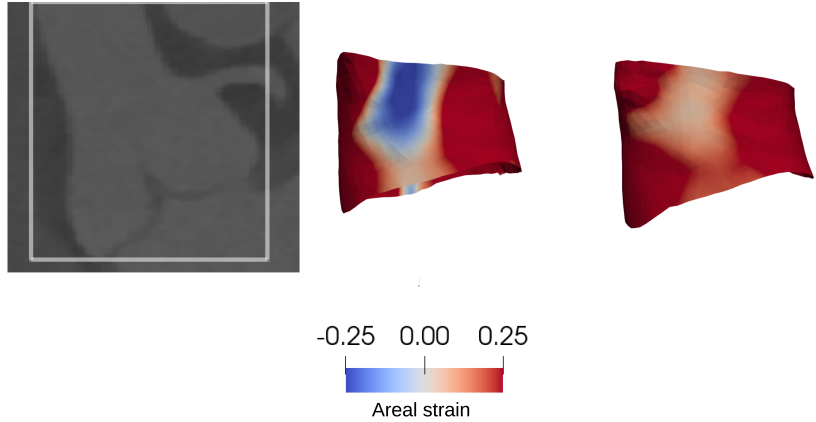}
    \caption{Spatial heterogeneity of areal strain among three subjects with high-grade AR. For each case, the CTA image is shown alongside strain maps from two consecutive end-systolic frames to illustrate the temporal consistency of the observed spatial pattern across adjacent frames. Subjects A (age 59) and B (age 50) demonstrate low median strain despite focal regions of elevated strain. Subject A had a root aneurysm and valvular calcification, while Subject B had visible valvular calcification. Conversely, Subject C (age 22) demonstrates elevated median strain with a focal low-strain region and no visible calcification. White boxes indicate the aortic region visualized in the strain maps, and arrows indicate visible calcification. These cases illustrate within-subject spatial strain patterns and are not intended for direct quantitative comparison of strain magnitude across subjects. The color bar is clipped at $\pm 0.25$ for visualization.}
    \label{fig:median_strain_across_ar_severity_groups}
\end{figure}

\paragraph{Strain-map patterns in high-grade aortic regurgitation}
Despite all three subjects having high-grade AR, their spatial strain patterns differed markedly (Fig.~\ref{fig:median_strain_across_ar_severity_groups}). Subject A (age 59), with an asymmetric aortic root aneurysm and valvular calcification, demonstrated low median and 95th-percentile strain (-0.05 and 0.06, respectively), despite a focal region of elevated strain. Subject B (age 50) also had valvular calcification and low median strain (-0.04) with a focal high-strain region, with a 95th-percentile strain of 0.17. Subject C (age 22), without visible calcification, demonstrated substantially higher median and 95th-percentile strain (0.36 and 0.92, respectively) than the other high-grade AR subjects.

These observations suggest that clinically defined AR severity may not fully capture biomechanical heterogeneity. Although the small number of high-grade AR subjects ($n = 3$) precludes meaningful statistical inference, these differences motivate further investigation of patient-specific factors, including age, calcification, and root geometry, that may contribute to variation in aortic strain patterns.

\section{Discussion}
\label{sec:discussion}
We introduced a predominantly automated framework for constructing ascending aortic areal strain maps from 4D CTA and characterizing strain accuracy and reliability across segmentation sources and spatial scales. Although image-based aortic strain is not yet a clinical metric, related biomechanical measures have shown prognostic value: FE–derived aortic wall stress has been associated with disease progression and adverse events \cite{zamirpour_temporal_2024}, and reduced ascending aortic distensibility has been linked to all-cause mortality and major cardiovascular events \cite{redheuil_proximal_2014}. Unlike model-derived wall stress and pressure-normalized measures such as distensibility, image-derived strain can be computed directly from wall deformation without explicit constitutive modeling or pressure normalization, making it attractive as an accessible marker of patient-specific mechanics. With modern GPU-accelerated registration methods \cite{jena_adaptive_2026}, time-resolved strain estimation may be computationally compatible with clinical workflows.

Our findings, based on analysis of end-systolic strain, help characterize the conditions under which image-derived strain can be interpreted as a reliable biomechanical marker. We used the FE-validation framework to inform DR parameter selection and establish DR-derived strain accuracy against FE-derived ground truth, and characterized the reliability of global strain measures across segmentation sources. Spectral analysis further showed that cross-segmentation reliability varied with spatial scale, with greater agreement for broad deformation patterns than for finer spatial features. The close agreement of median strain measures across segmentation sources supports the integration of automated segmentation into strain pipelines, while the spatial analysis characterizes how cross-segmentation reliability varies with spatial scale.

\subsubsection{Strain accuracy and reliability}

FE simulations provided both a reference standard for registration parameter selection and a controlled framework for evaluating the accuracy of registration-derived strain. Across FE-validation subjects, registration-derived median strain showed minimal bias relative to the known FE reference for both observer-derived meshes (bias = 0.0013; 95\% LoA, $[-0.0023,\,0.0049]$) and nnU-Net-derived meshes (bias = 0.0056; 95\% LoA, $[-0.0068,\,0.0179]$), indicating little systematic over- or underestimation within the deformation regime studied. For 95th-percentile strain, observer-derived meshes exhibited a larger positive bias than nnU-Net-derived meshes (0.0108 versus 0.0038), whereas nnU-Net-derived meshes showed wider LoA ($[-0.0265,\,0.0340]$ versus $[0.0022,\,0.0194]$), indicating greater variability relative to the FE reference despite lower systematic bias.

Separately, we compared registration-derived strain across segmentation sources using Bland–Altman analysis. Median end-systolic areal strain showed minimal bias and narrow LoA for both the interobserver (bias = -0.0045; 95\% LoA, $[-0.0490,\,0.0399]$) and Observer-1–nnU-Net (bias = -0.0052; 95\% LoA, $[-0.0340,\,0.0237]$) comparisons. Differences were more pronounced for 95th-percentile strain: the interobserver comparison showed greater negative bias and substantially wider LoA (bias = -0.0371; 95\% LoA, $[-0.1810,\,0.1068]$), whereas the Observer-1–nnU-Net comparison showed minimal bias and narrower LoA (bias = 0.0029; 95\% LoA, $[-0.0347,\,0.0406]$). Median areal strain was highly reproducible across segmentation sources, whereas 95th-percentile strain showed greater sensitivity to segmentation variability, particularly in the interobserver comparison. The closer agreement in 95th-percentile strain observed between Observer-1– and nnU-Net-derived strain should be interpreted in the context of the nnU-Net model having been trained on Observer-1-derived segmentations.

Together, the FE-validation and cross-segmentation analyses support FE-informed parameter selection as an important component of registration-derived strain estimation. Optimal registration settings may depend on both segmentation source and imaging characteristics; observer-derived and nnU-Net-derived meshes exhibited distinct but partially overlapping low-loss parameter regimes (Fig.~\ref{fig:dr_parameter_tuning_ncc3x3}). Rather than assuming a single universal registration configuration, future registration-derived strain pipelines may benefit from task-specific parameter tuning against controlled biomechanical references. Additionally, the close agreement of median strain across segmentation sources supports the incorporation of automated segmentation into strain pipelines.

\subsubsection{Scale-dependent strain reliability}

Spectral analysis enabled us to quantify strain reliability across spatial scales using LB modes, which are derived from the surface geometry and provide a basis for separating broad spatial patterns from progressively finer structure. Cross-segmentation strain deviation, normalized by the raw deviation, was lowest when strain fields were reconstructed using low-frequency modes and increased as higher-frequency modes were progressively restored. Thus, large-scale deformation patterns remained stable across segmentation sources, whereas disagreement increased at finer spatial structure. This suggests that low-frequency strain features are less sensitive than higher-frequency features to small segmentation differences. These findings may help define the spatial scales at which regional strain patterns can be interpreted reliably in clinical applications.

\subsubsection{Comparison with previous strain studies}

Direct comparison of absolute strain magnitudes with prior studies is challenging because aortic strain has been quantified using different strain definitions and spatial representations. Prior studies have measured circumferential strain within individual two-dimensional cross-sections using PC-MRI \cite{wedding_measurement_2002, draney_quantification_2002}, assessed spatially resolved circumferential deformation within individual aortic cross-sections using CMR \cite{wilson_assessment_2019, haraldsson_feasibility_2014}, or speckle-tracking echocardiography \cite{teixeira_circumferential_2013}, and estimated three-dimensional circumferential and longitudinal strain from CTA \cite{huckaby_wall_2025}. Others have quantified cross-sectional area strain from the maximum and minimum luminal areas over the cardiac cycle \cite{hrabak-paar_variability_2020}. Differences in strain definition, dimensionality, and spatial resolution limit direct numerical comparison of reported magnitudes.

Despite these methodological differences, previous work consistently demonstrates that aortic deformation is spatially heterogeneous rather than uniform \cite{draney_quantification_2002, haraldsson_feasibility_2014, huckaby_wall_2025, draney_quantification_2004}, consistent with the regional strain patterns observed in our cohort. Karatolios et al. reported that two abdominal aortic aneurysm patients exhibited very low mean circumferential strain ($<1\%$) despite localized peak strains of 12–18\%, together with regions of negative strain \cite{karatolios_method_2013}. Similarly, the aneurysm case in our cohort demonstrated low overall strain with localized regions of elevated strain interspersed with areas of negative strain. Together, these findings illustrate how global strain measures can obscure substantial regional variation.

\subsubsection{Clinical relevance and future directions}
Maximum ascending aortic diameter is an established marker of aortic enlargement used to guide surveillance and prophylactic surgical thresholds \cite{isselbacher_2022_2022}. The observed inverse associations between 95th-percentile areal strain and diameter, AHI, and age are consistent with prior reports linking larger aortic diameter and increased age to reduced strain \cite{bieseviciene_two-dimensional_2017, h_evaluation_2022, redheuil_reduced_2010}. However, the moderate strength of these associations suggests that these clinical variables do not fully explain strain variability, motivating future investigation of localized strain features and additional contributors to aortic mechanical heterogeneity.

The qualitative AR analysis suggests that strain maps may capture biomechanical heterogeneity not reflected by clinical AR severity alone. Among the three high-grade AR subjects, strain patterns ranged from diffuse elevated strain with focal low-strain regions to low median strain with focal high-strain regions despite similar clinical severity classifications. These differences may reflect patient-specific variation in age, valve morphology, aortic geometry, and calcification. The elevated median strain observed in the youngest high-grade AR subject is consistent with prior work reporting greater normalized aortic root volume change in BAV patients with moderate-to-severe AR \cite{pouch_dynamic_2021}, whereas the lower median strain observed in the two older subjects with calcific valve changes is consistent with literature linking aging and calcification to reduced aortic compliance \cite{redheuil_reduced_2010, mceniery_aortic_2009, cohoon_relationship_2018}. Future work should evaluate these findings in larger cohorts to determine whether distinct strain phenotypes represent reproducible markers of aortic remodeling.

\subsubsection{Image-derived strain as an input to patient-specific computational modeling}

Patient-specific aortic simulations have been used both to infer otherwise inaccessible mechanical properties and to predict the mechanical consequences of intervention. The accuracy of these simulations can depend on inputs that are difficult to obtain in vivo, including the unloaded configuration, regional material properties, residual stress, and wall thickness. Excised surgical tissue can provide direct experimental measurements of some of these quantities, particularly tissue material behavior and wall thickness \cite{wisneski_bicuspid_2015}, but it is available only after intervention and therefore cannot generally support preoperative planning. Dense image-derived strain maps provide spatially resolved in vivo observations of aortic deformation under physiological loading. These measurements could be used to calibrate constitutive parameters, constrain inverse estimation of the unloaded configuration and prestress state, and evaluate whether simulated deformation is consistent with observed patient-specific motion.

Image-derived deformation has been incorporated directly into inverse-modeling frameworks to personalize aortic simulations. Geronzi et al. used MRI-derived aortic motion to calibrate mechanical boundary-condition parameters in a structural finite-element model, improving agreement between simulated and observed patient-specific kinematics \cite{geronzi_calibration_2023}. Latorre et al. similarly used MRI-derived displacement for patient-specific inverse FE modeling; importantly, separate synthetic validation experiments showed that principal-strain objectives recovered prescribed constitutive behavior more accurately than displacement-based objectives \cite{latorre_molins_estimating_2025}. These studies establish the feasibility of using image-derived deformation to constrain patient-specific aortic models, while the synthetic findings of Latorre et al. suggest that reliable strain measurements could provide particularly informative constraints for future inverse-modeling applications.

Patient-specific simulation of thoracic endovascular aortic repair (TEVAR) remains sparse and has largely been demonstrated in proof-of-concept studies \cite{romarowski_computational_2018,derycke_computational_2024}. Romarowski et al. virtually compared three endograft configurations for treatment of an ascending-aortic pseudoaneurysm, identifying a custom configuration with improved predicted wall apposition, although the vessel wall was modeled as rigid and the proposed intervention was not performed clinically \cite{romarowski_computational_2018}. In an adjacent application involving type B aortic dissection, Kan et al. demonstrated clinically anchored validation of a patient-specific deployment simulation against follow-up imaging, although aortic wall mechanics were represented using population-derived rather than patient-specific material properties \cite{kan_patient-specific_2021}. 

Longitudinal imaging provides an opportunity to characterize patient-specific mechanical changes following intervention.
Tjahjadi et al. applied DR to ECG-gated CTA images to quantify ascending aortic motion, cross-sectional area strain, and three-dimensional growth in patients undergoing zone 2/3 TEVAR. They observed reduced ascending aortic motion and increased aortic strain post-TEVAR, noting that lower post-TEVAR strain was associated with higher aortic growth rates \cite{tjahjadi_three-dimensional_2025}. These findings suggest that image-derived strain may help identify patients at risk of progressive post-TEVAR ascending aortic dilation and could help provide patient-specific mechanical information for simulations aimed at predicting postoperative changes, although this application has not yet been demonstrated.

Patient-specific modeling of vascular growth and remodeling may additionally help predict ascending thoracic aortic aneurysm progression. Mousavi et al. used a constrained-mixture FE model to simulate how localized elastin degradation could alter wall-specific mechanics in a patient-specific ascending thoracic aortic aneurysm \cite{mousavi_patient-specific_2019}. Because these predictions depend on assumed material behavior and remodeling parameters, longitudinal image-derived strain measurements could provide patient-specific observations for calibrating model parameters and evaluating predicted changes in wall mechanics or aortic geometry over time.

\subsubsection{A staged validation pathway toward clinical validity and utility}

Evaluation of dense strain estimates should be matched to the intended context of use and should distinguish controlled verification from clinical validity. Computationally prescribed deformation enables direct evaluation of registration-derived strain against dense, known ground truth for the controlled recovery task. De Craene et al. evaluated Temporal Diffeomorphic Free-Form Deformation (TDFFD) for estimating left-ventricular motion and strain from three-dimensional echocardiography across three distinct evidentiary levels: recovery of known motion in synthetic ultrasound sequences, physiological plausibility of segmental longitudinal-strain curves in healthy volunteers, and correspondence between changes in septal peak systolic strain and subsequent reverse remodeling in patients undergoing cardiac resynchronization therapy \cite{de_craene_temporal_2012}.

Our study addresses controlled verification and measurement reproducibility as foundations for subsequent assessment of clinical validity. FE-controlled deformation enables verification of recovery of known three-dimensional displacement and strain fields, and provides a scalable framework for evaluating variation across patient anatomies, acquisition conditions, and deformation states. Comparisons across observer- and deep-learning-derived segmentations quantify reproducibility as a function of Laplace–Beltrami spatial scale. Together, these analyses separate registration-recovery error from segmentation-induced variability and identify the spatial information sufficiently robust to support subsequent biological and clinical investigation.

The next stage should establish external reproducibility and clinical and prognostic validity in human cohorts. Longitudinal studies in BAV patients should determine whether image-derived regional strain provides information beyond conventional measures of aortic disease, predicts adverse remodeling or clinical events, and improves patient-specific risk stratification and treatment planning.

\subsubsection{Limitations and Assumptions}

Several limitations should be considered when interpreting these results. First, the CTA field of view was not standardized across subjects, and some acquisitions provided incomplete coverage of the ascending aorta. This real-world imaging heterogeneity may have contributed to variability in segmentation, registration, and the resulting strain maps. Second, FE simulations used homogeneous, literature-defined material properties and simplified boundary conditions, without patient-specific material calibration or fiber-orientation fields. They were intended as controlled references for DR parameter optimization rather than patient-specific biomechanical models. Finally, the cohort was relatively small, with only three high-grade AR subjects, limiting our ability to link specific strain patterns to pathological changes; the AR analysis should therefore be interpreted as hypothesis-generating.

\section{Conclusion}
\label{sec:conclusion}
This study establishes a predominantly automated framework for estimating ascending aortic areal strain from 4D cardiac CTA and evaluating its accuracy and reliability from global summaries to spatially resolved strain patterns. By combining FE-based accuracy assessment with cross-segmentation comparisons and spectral reliability analysis, the framework provides a basis for determining which features of image-derived strain can be interpreted reliably. Future validation in larger clinical cohorts will be needed to establish whether regional strain provides prognostic information beyond conventional measures of aortic disease.

\section*{CRediT authorship contribution statement}
\label{sec:contribution}
\textbf{Tricia Lobo:} Conceptualization, Investigation, Methodology, Software, Validation, Formal analysis, Visualization, Writing -- Original Draft, Writing -- Review and Editing, Data Curation, Project Administration. 
\textbf{Wensi Wu:} Conceptualization, Methodology, Software, Writing -- Review \& Editing. 
\textbf{Harold I. Litt:} Conceptualization, Investigation, Methodology, Resources, Funding acquisition, Supervision, Writing -- Review \& Editing.
\textbf{Nimesh D. Desai:} Conceptualization, Investigation, Methodology, Resources, Funding acquisition, Supervision.
\textbf{Alison M. Pouch:} Conceptualization, Methodology, Supervision, Resources, Funding acquisition, Writing -- Review \& Editing.

\section*{Acknowledgements}
\label{sec:acknowledgements}
The authors thank Shir Goldfinger for assistance with data curation and dataset organization, and Jilei Hao for development of 4D image manipulation and I/O functionality in ITK-SNAP used to support image analysis in this study. The authors thank the clinical research professionals at Penn RADCORE for imaging research support. They appreciate the clinical insight and resources provided by Drs. Joseph Bavaria, Victor Ferrari, and Thomas Gleason for 4D CT imaging. The authors also thank Paris Perdikaris for valuable discussions on Laplace–Beltrami methods and Elaine Tseng for clinical insights on strain map reproducibility.

\section*{Funding sources}
\label{sec:funding}
This work was supported by the National Heart Lung and Blood Institute (R01-HL163202) and the Dean's Innovation Fund at the University of Pennsylvania. WW acknowledges support from the National Heart Lung and Blood Institute (K25-HL168235) and the Additional Ventures Single Ventricle Research Fund (1434814).   

\newpage
\printbibliography
\newpage

\clearpage
\appendix
\setcounter{figure}{0}
\renewcommand{\thefigure}{A\arabic{figure}}
\setcounter{table}{0}
\renewcommand{\thetable}{A\arabic{table}}
\section{Appendix}
\label{sec:appendix}
\subsection{Full cohort characteristics}

The primary validation analyses were performed using a curated 14-patient subset to avoid overlap with segmentation model development. Exploratory clinical association analyses were performed using all 31 eligible 4D CTA cases. Demographic and clinical characteristics of the full eligible cohort are summarized in Table~\ref{tab:full_cohort_characteristics}.

\begin{table}[htbp]
\centering
\caption{Full Cohort Characteristics}
\label{tab:full_cohort_characteristics}
\begin{tabular}{l c}
\hline
\textbf{Characteristic} & \textbf{Value} \\
\hline
Total patients, n & 31 \\
Age, y & 50.0 (41.5--61.0) \\
Female sex, \% & 22.6 \\
Height, m & 1.78 (1.70--1.83) \\
Body mass index, kg/m$^2$ & 28.1 (25.7--30.6) \\
Ascending aortic diameter, cm & 4.4 (3.9--4.6) \\
Height-indexed aortic diameter, cm/m & 2.5 (2.2--2.6) \\
Left ventricular ejection fraction, \% & 58 (56--63) \\
$>$ Mild aortic regurgitation, \% & 58.1 \\
Follow-up visits, n & 1.0 (0--2) \\
Patients with $\geq$1 follow-up, \% & 64.5 \\
\addlinespace
\multicolumn{2}{l}{Aortic valve morphology, \%} \\

\quad L/R fusion & 58.1 \\
\quad R/N fusion & 19.4 \\
\quad Partial L/R fusion & 9.7 \\
\quad Trileaflet & 6.5 \\
\quad Dysmorphic morphology & 3.2 \\
\quad Sievers type 0 bicuspid valve, AP orientation & 3.2 \\
\hline
\end{tabular}
\smallskip

\parbox{0.60\textwidth}{%
\footnotesize
Values are reported as median (IQR), percentage, or count, as appropriate.
L/R, left--right coronary cusp fusion; R/N, right--noncoronary cusp fusion;
AP, anterior--posterior. Dysmorphic morphology denotes abnormal leaflet anatomy
that could not be clearly classified as Sievers type 0 or type 1. Sievers type 0
denotes a bicuspid valve with two symmetric cusps and no raphe.
}
\end{table}

\subsection{Supplementary FE validation}

\begin{figure}[ht!]
  \centering
  \includegraphics[width=\textwidth]{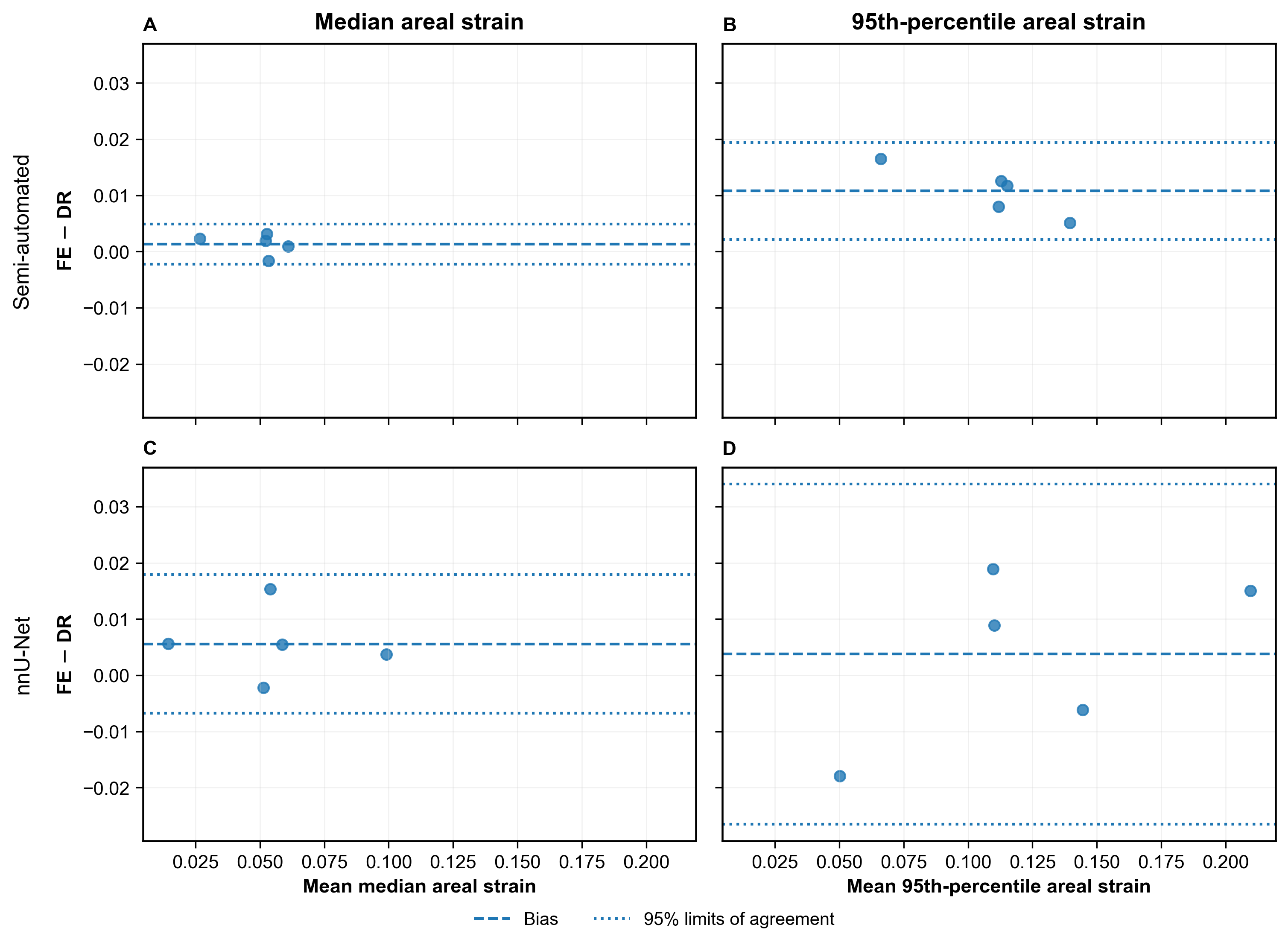}
    \caption{
Bland--Altman analysis of DR-derived areal strain versus FE reference for the five FE-validated subjects. Analysis was performed using the selected $\alpha$ and $\beta$ registration parameters. Panels (A) and (B) show Bland-Altman analysis of median and 95th percentile strain for semi-automated segmentations, respectively. Panels (C) and (D) show the corresponding analyses for nnU-Net-derived segmentations. Dashed lines indicate the mean bias and dotted lines denote the 95\% limits of agreement.} \label{fig:bland_altman_combined_analysis}
\end{figure}

Across the five FE-validated subjects, Bland--Altman analysis demonstrated small signed bias for both segmentation sources (Table~\ref{tab:fe_dr_comparison}, Figure~\ref{fig:bland_altman_combined_analysis}). Bias remained small relative to the overall strain range, although limits of agreement were wider for the 95th-percentile strain metric than for median strain.


\end{document}